\documentclass[times]{aastex7}
\usepackage{dcolumn}
\usepackage{bm}
\usepackage{longtable}
\usepackage{mathrsfs}
\usepackage{graphicx,epsfig,latexsym,amssymb}
\usepackage{multirow,amsmath,array,booktabs,color}
\usepackage[section]{placeins}
\usepackage{soul}

\setcitestyle{numbers}

\graphicspath{{./}}

\makeatletter
\def\frontmatter@above@affiliation@script{%
 \skip@\@flushglue
 \@flushglue\z@ plus.3\hsize\relax
 \centering
 \@flushglue\skip@
 \addvspace{15\p@}%  %% was 3.5\p@ — increased to match collaboration-gap height
}%
\makeatother
\usepackage{ulem}
\begin{document}

\title{Multi-Model Analysis of the Astrophysical $S(E)$ Factor for the $^{9}\mathrm{Be}(p,\alpha)^{6}\mathrm{Li}$ Reaction and Its Impact on Astrophysical Reaction Rates}

\author{Xue-Jian Wang}
\affiliation{School of Physics, Anhui University, Hefei 230601, China}
\email{}

\author{Wei-Ke Nan}
\affiliation{School of Physics, Southern University of Science and Technology, Shenzhen 518055, China}
\email{}
\author{Qing Wang}
\affiliation{School of Physics, Southern University of Science and Technology, Shenzhen 518055, China}
\affiliation{College of Physics and Optoelectronic Engineering, Shenzhen University, Shenzhen 518000, China}
\email{}

\author{Tian-Yu Tao}
\affiliation{School of Physics, Anhui University, Hefei 230601, China}
\email{}

\author{Qun-Gang Wen}
\affiliation{School of Physics, Anhui University, Hefei 230601, China}
\email[show]{\parbox[t]{0.12\textwidth}{}
\parbox[t]{0.8\textwidth}{
    qungang@ahu.edu.cn \\
    jianyou@ahu.edu.cn \\
    lichengbo@bjast.ac.cn \\
    jiahm@cnncmail.cn}}

\author{Jian-You Guo}
\affiliation{School of Physics, Anhui University, Hefei 230601, China}
\email{}

\author{Cheng-Bo Li}
\affiliation{Institute of Radiation Technology, Beijing Academy of Science and Technology, Beijing 100089, China}
\email{}

\author{Hui-Ming Jia}
\affiliation{Department of Nuclear Physics, China Institute of Atomic Energy, Beijing 102413, China}
\email{}

%% Use the \collaboration command to identify collaborations. This command
%% takes an optional argument that is either a number or the word "all"
%% which tells the compiler how many of the authors above the command to
%% show. For example "\collaboration[all]{(DELVE Collaboration)}" wil include
%% all the authors above this command.

%% Mark off the abstract in the ``abstract'' environment.
\begin{abstract}

The $^{9}\mathrm{Be}(p,\alpha)^{6}\mathrm{Li}$ reaction is a destruction process for $^{9}\text{Be}$ in stars and the Big Bang. The Trojan Horse Method (THM) is a well-established indirect technique that enables the determination of reaction cross sections within the Gamow energy region while avoiding the uncertainties associated with low-energy extrapolations of the $S(E)$ factor and electron-screening effects. In this work, the bare-nucleus THM data reported by Wen \textit{et al.} are systematically analyzed using polynomial fitting, a double Breit--Wigner model, and the $R$-matrix formalism. The applicability and physical implications of these theoretical approaches in describing the low-energy $S(E)$ factor of the $^{9}\mathrm{Be}(p,\alpha)^{6}\mathrm{Li}$ reaction are investigated and compared. Since the THM data have been incorporated into the NACRE II reaction-rate compilation, leading to improved accuracy in reaction-rate evaluations, the refined theoretical analysis presented here provides further insight into the underlying reaction dynamics and establishes a more reliable foundation for stellar reaction-rate calculations. The results obtained in this work offer valuable reference for future high-precision experimental investigations and the development of theoretical models in nuclear astrophysics.

\end{abstract}

%% Keywords should appear after the \end{abstract} command.
%% The AAS Journals now uses Unified Astronomy Thesaurus (UAT) concepts:
%% https://astrothesaurus.org
%% You will be asked to selected these concepts during the submission process
%% but this old "keyword" functionality is maintained in case authors want
%% to include these concepts in their preprints.
%%
%% You can use the \uat command to link your UAT concepts back its source.

%% From the front matter, we move on to the body of the paper.
%% Sections are demarcated by \section and \subsection, respectively.
%% Observe the use of the LaTeX \label
%% command after the \subsection to give a symbolic KEY to the
%% subsection for cross-referencing in a \ref command.
%% You can use LaTeX's \ref and \label commands to keep track of
%% cross-references to sections, equations, tables, and figures.
%% That way, if you change the order of any elements, LaTeX will
%% automatically renumber them.

\section{Introduction}

Nuclear reactions in astrophysical environments govern the synthesis of elements and serve as a fundamental physical basis for understanding the origin and evolution of the Universe. The abundances of the light elements—lithium (Li), beryllium (Be), and boron (B)—are of particular significance in nuclear astrophysics~\cite{Tumino2025,Rolfs1988}. Although approximately 10\% of lithium is synthesized during Big Bang nucleosynthesis (BBN), lithium, beryllium, and boron are predominantly produced through cosmic-ray spallation of carbon (C), nitrogen (N), and oxygen (O) nuclei~\cite{Vangioni2000}. The primordial abundance of lithium serves as a key observable for testing BBN models. The $^{9}\mathrm{Be}(p,\alpha)^{6}\mathrm{Li}$ reaction is of limited relevance to BBN, becoming significant only in neutron rich regions of inhomogeneous BBN (IBBN) models. By contrast, its principal astrophysical role is the destruction of $^{9}\mathrm{Be}$ in stellar interiors, which provides the primary astrophysical motivation for the present study~\cite{Kajino1990,Boyd1989}.

In stellar interiors, the depletion of ${}^{9}\text{Be}$ is primarily driven by proton-induced $(p, \alpha)$ reactions, with the ${}^{9}\text{Be}(p, \alpha){}^{6}\text{Li}$ reaction cross-section in the stellar energy range of $10\text{--}100$ keV being particularly critical. In this energy regime, which lies well below the Coulomb barrier, the reaction cross-section decreases exponentially as the incident energy decreases \cite{Lamia2015, Sierk1973}. Such extremely low reaction probabilities present significant challenges for direct experimental measurements. To facilitate physical analysis and reliable low-energy extrapolation of experimental data, the cross-section is typically described by the astrophysical $S(E)$ factor, defined as:
\begin{equation}
    \sigma(E) = \frac{S(E)}{E} \exp(-2\pi\eta)
\end{equation}
where $E$ is the center-of-mass energy and $\eta$ is the Sommerfeld parameter. The primary exit channels and corresponding $Q$-values for the ${}^{9}\text{Be}+p$ reaction are:
\begin{align}
    {}^{9}\text{Be} + p &\rightarrow {}^{6}\text{Li} + \alpha \quad (Q=2.13 \text{ MeV}) \\
    {}^{9}\text{Be} + p &\rightarrow {}^{8}\text{Be} + d \quad (Q=0.56 \text{ MeV}) \\
    {}^{9}\text{Be} + p &\rightarrow {}^{10}\text{B} + \gamma \quad (Q=6.59 \text{ MeV})
\end{align}

Research on this reaction has been extensive and complex. In 1973, Sierk and Tombrello performed the first R-matrix analysis using angular distribution and cross-section data in the range of $30\text{--}700$ keV, incorporating a sub-threshold resonance state \cite{Sierk1973}. Although this model successfully reproduced the overall structure of the experimental data, significant uncertainties in the low-energy region (approximately 55\%), primarily limited by the counting statistics and background suppression techniques of that era, led to ongoing controversies regarding the true contribution of the sub-threshold resonance.

Subsequently, in 1997, Zahnow et al. extended the measurement range down to $16$ keV \cite{Zahnow1997}. By applying a Breit-Wigner formula with explicit interference terms, they observed a significant enhancement in the astrophysical $S(E)$ factor at low energies. They argued that this enhancement could not be fully accounted for by the sub-threshold resonance alone, and subsequently extracted an anomalous electron screening potential of $U_e \approx 900$ eV, which triggered extensive discourse on the fundamental reaction mechanisms at sub-Coulomb barrier energies. Following these experimental developments, Barker and Kondō \cite{Barker2001} refined the description of the ${}^{10}\text{B}$ compound nucleus level structure in 2001, employing a multi-level R-matrix framework to provide a more rigorous theoretical treatment of the resonance states.

To bypass the suppression of the Coulomb barrier and the impact of electron screening inherent in direct measurements, the Trojan Horse Method (THM) was introduced. Studies by Lamia et al.~\cite{Lamia2012, Lamia2013} and subsequently Wen et al.~\cite{Wen2008} utilized quasi-free reaction mechanisms to successfully extract bare nuclear cross-sections. This approach eliminates the need for low-energy extrapolation and successfully reproduces the resonance peak at $E_{c.m.} \approx 0.27$ MeV. However, discrepancies persist between the $S(0)$ and $U_e$ values derived from THM and those obtained via traditional direct measurements. In 2018, Fang et al.~\cite{Fang2015, Fang2016, Fang2018} conducted high-density measurements in the energy range of $18\text{--}100$ keV using the thick-target yield method. By refining the experimental data processing workflow, they obtained more rigorous $U_e$ values, which provide significantly stronger constraints for quantitative evaluations within this energy regime.

Despite the accumulation of increasingly precise experimental data, divergent interpretations persist within current theoretical frameworks. Various research groups have employed different theoretical approaches, ranging from multi-level $R$-matrix analyses~\cite{Sierk1973,Barker2001} to interference-corrected Breit--Wigner models~\cite{Zahnow1997,Fang2018}. These approaches have resulted in significant disparities in the extracted contributions of sub-threshold resonances and non-resonant backgrounds. Such inconsistencies inevitably result in model-dependent extrapolations of the astrophysical $S(E)$ factor. Consequently, a systematic comparison of the fitting characteristics of these theoretical frameworks, aimed at elucidating the physical nature of the low-energy cross-section enhancement—whether driven by sub-threshold resonances or coherent interference—has become a pivotal step toward a deeper understanding of nuclear reaction mechanisms at sub-barrier energies. In this paper, we systematically analyze the cross-section enhancement behavior by integrating existing direct measurement datasets with THM-derived bare nuclear benchmark data, employing a multi-theoretical model fitting approach to critically evaluate and dissect these fundamental scientific issues.

\section{Theoretical Framework}

\subsection{Interference-corrected Breit-Wigner Method}

The total astrophysical $S_{\text{bare}}(E)$ factor for bare nuclei, as proposed by Zahnow et al. \cite{Zahnow1997, Fang2018}, is decomposed into three components: a resonant contribution from the $J^{\pi} = 1^{-}$ state at $E_{\text{lab}} = 336$ keV, a direct process term, and an interference term. The expression is given by:
\begin{equation}
\begin{split}
S_{bare}(E) = & S_R(E_p) + S_D(E_p) \\
              & + 2\sqrt{S_R(E_p) X_D S_D(E_p)} \cos(\delta_R)
\end{split}
\end{equation}
where $S_R(E_p)$ is the resonance term described by the single-level Breit-Wigner formula, incorporating a fixed $\alpha$-particle partial width $\Gamma_{\alpha}$, and energy-dependent proton partial width $\Gamma_p(E_p)$ and total width $\Gamma(E_p)$. The term $S_D(E_p)$ represents the direct process, which is generally assumed to be energy-independent \cite{Zahnow1997}. Within the interference term, $X_D$ denotes the fraction of the direct (non-resonant) cross-section that interferes with the resonance, and $\delta_R$ represents the resonance phase shift, defined as:
\begin{equation}
\delta_R(E_p) = \arctan\left( \frac{\Gamma_R}{2(E_p - E_R)} \right).
\label{eq:phase_shift}
\end{equation}

\subsection{Double Breit-Wigner Framework}

The resonance structures above and below the threshold, characterized by distinct spin-parity ($J^{\pi}$) and orbital angular momentum ($l$) channels, possess partial wave amplitudes that are mutually orthogonal. Upon integrating the angular distributions over the full solid angle to obtain the total cross-section, the interference terms between different partial waves vanish. Consequently, their contributions to the low-energy astrophysical $S(E)$ factor can be treated incoherently—that is, interference effects are neglected, and contributions are linearly summed. This approach is not only consistent with the principles of angular momentum coupling but also facilitates the evaluation of the relative contribution of each resonant structure to the astrophysical factor in the low-energy regime \cite{Wen2008, Wen2016, Wang2024}.

To further investigate the impact of the sub-threshold resonance on the astrophysical $S(E)$ factor, a double Breit-Wigner model is utilized to fit the THM data. The total astrophysical $S$-factor is expressed as:
\begin{equation}
S(E) = S_{0}(E) + S_{1}(E), \label{eq:total_S}
\end{equation}
where $S_{0}(E)$ and $S_{1}(E)$ are the individual resonance terms, formulated as:
\begin{equation}
S_{0}(E) = \frac{A_{0} \cdot (\Gamma_{0}/2)^{2}}{(E - E_{0})^{2} + (\Gamma_{0}/2)^{2}}, \label{eq:S0}
\end{equation}
\begin{equation}
S_{1}(E) = \frac{A_{1} \cdot (\Gamma_{1}/2)^{2}}{(E - E_{1})^{2} + (\Gamma_{1}/2)^{2}}. \label{eq:S1}
\end{equation}
Here, $S_{0}(E)$ describes the resonance peak above the threshold, while $S_{1}(E)$ accounts for the contribution of the sub-threshold resonance with a resonance energy of $E_{1} = -23$ keV. In this model, $S_{0}(0)$ denotes the astrophysical $S(E)$ factor at $E=0$ derived from a single-resonance fit, whereas $S(0)$ represents the total astrophysical $S(E)$ factor obtained by incorporating the sub-threshold resonance (the double-resonance model).

\subsection{R-matrix Theory}

As a core framework for describing resonance behavior and cross-section extrapolation in nuclear reaction physics, R-matrix theory is recognized as one of the most rigorous theoretical tools due to its systematic and self-consistent treatment of resonant states, interference effects, and background contributions. Since its proposal by Wigner and Eisenbud in 1948 and subsequent expansion by Lane and Thomas, it has become the standard method for analyzing nuclear reaction data \cite{Lane1958, Sierk1973}. Furthermore, the extension of its application to $\beta$-decay and radiative capture processes by Barker et al. \cite{Barker2001}, combined with the parametrization method introduced by Brune, has enabled R-matrix theory to integrate more closely with experimental data, establishing a highly refined theoretical system.

R-matrix theory demonstrates significant physical advantages in handling low-energy nuclear reactions \cite{Fang2016}. Particularly in complex systems where interference effects are prominent, single-resonance models often fail to describe the cross-section accurately due to their inability to account for multi-level interference. In contrast, R-matrix theory precisely characterizes the evolution of the cross-section by parameterizing the properties of compound nucleus levels \cite{Zahnow1997}. With the maturity of computational codes such as AZURE2, this theory has become a powerful tool for analyzing reaction cross-sections and data extrapolation in the nuclear astrophysics energy regime \cite{Fang2018}. In this study, we systematically compare the R-matrix fitting results generated by the AZURE2 code with those of the Breit-Wigner models to explore the intrinsic differences in the cross-section enhancement mechanisms under different physical frameworks.

\subsection{Trojan Horse Method}

\begin{figure}[htbp]
\centering
\includegraphics[width=10cm]{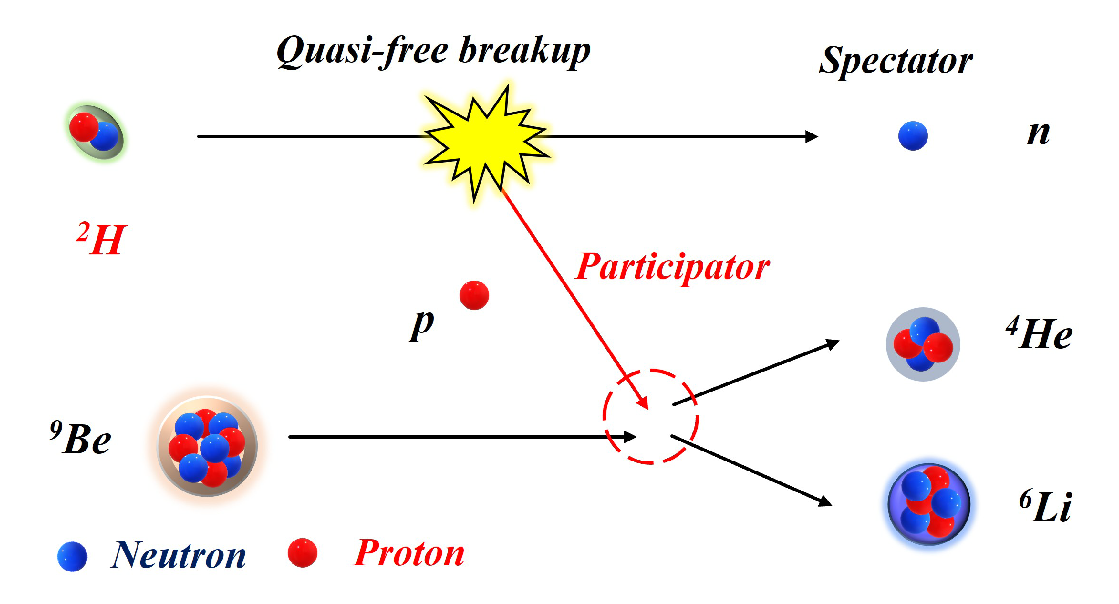}
\caption{(Color online) Schematic diagram of the quasi-free three-body reaction process with ${}^{2}\text{H}$ as the Trojan Horse nucleus.}
\label{wi}
\end{figure}

The Trojan Horse Method is a powerful indirect measurement technique in experimental nuclear physics, designed to investigate reaction mechanisms that are otherwise inaccessible at ultra-low energies \cite{Tumino2025}. Based on the quasi-free (QF) reaction mechanism, this method facilitates the extraction of the astrophysical $S(E)$ factor for two-body reactions by selecting appropriate three-body reaction processes, thereby effectively bypassing the Coulomb barrier suppression and the environmental electron screening effects \cite{Lamia2012, Spitaleri2014, Wen2011, Li2017}. While this section provides a concise overview of the method, its robust theoretical foundations have been extensively validated in numerous astrophysical studies \cite{Lamia2013}.

Figure~\ref{wi} illustrates the QF three-body reaction process using ${}^{2}\text{H}$ as the Trojan Horse nucleus. In the case of the ${}^{9}\text{Be}(p, \alpha){}^{6}\text{Li}$ reaction, the deuteron (${}^{2}\text{H} = p + n$) serves as the Trojan Horse. Within the QF mechanism, the proton ($p$) acts as the active participant in the reaction, while the neutron ($n$) remains as a spectator, preserving its initial momentum distribution (the Fermi motion) inside the deuteron. By kinematically selecting the QF events where the spectator momentum is approximately zero, the bare nuclear cross-section is extracted directly.

\section{Analysis and Discussion}

\subsection{Reaction cross sections and astrophysical factor S(E)}

\begin{figure}[htbp]
\centering
\includegraphics[width=10cm]{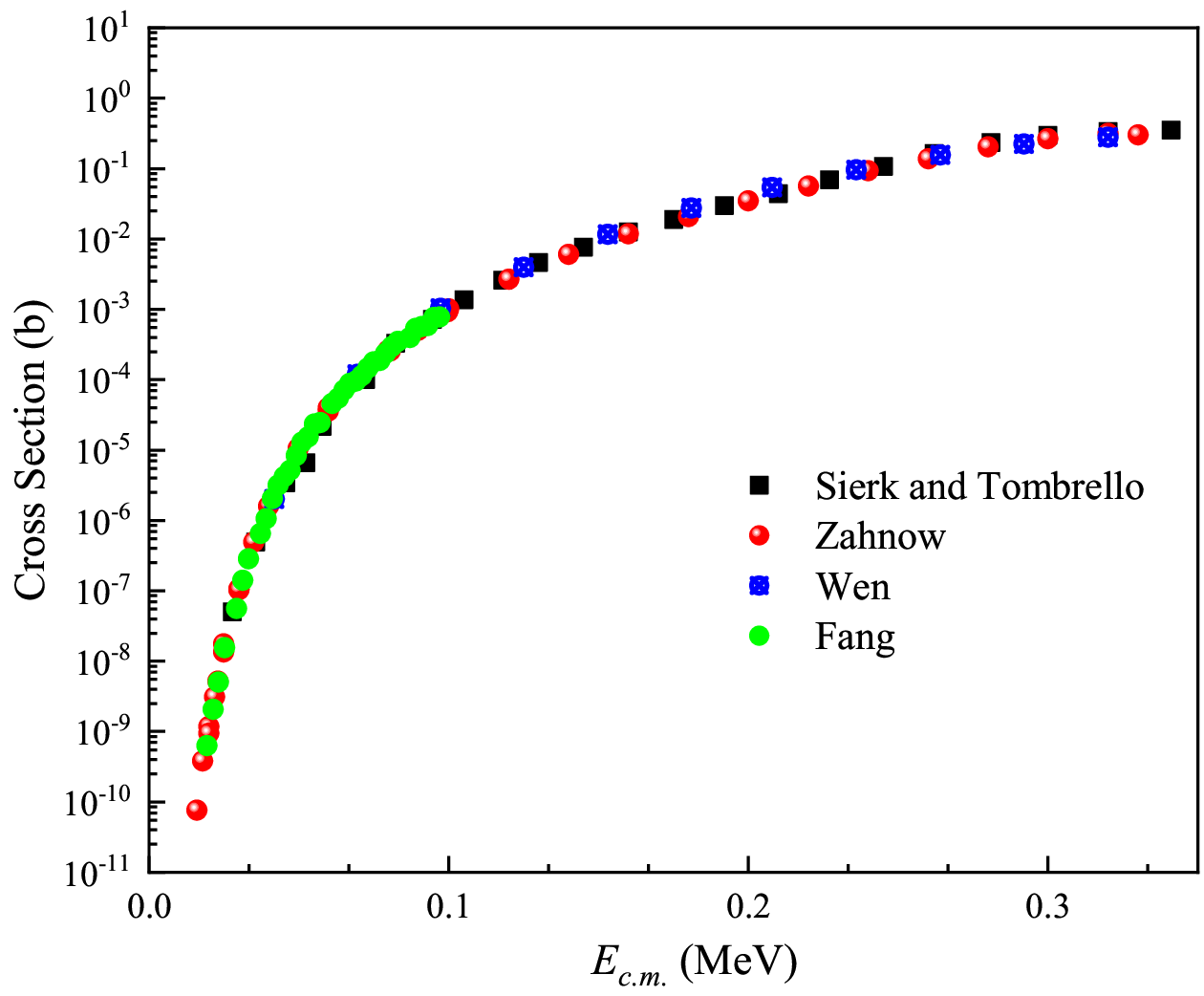}
\caption{(Color online) Cross section for the ${}^{9}\text{Be}(p, \alpha){}^{6}\text{Li}$ reaction as a function of center-of-mass energy ($E_{\text{c.m.}}$).}
\label{wii}
\end{figure}

\begin{figure}[htbp]
\centering
\includegraphics[width=10cm]{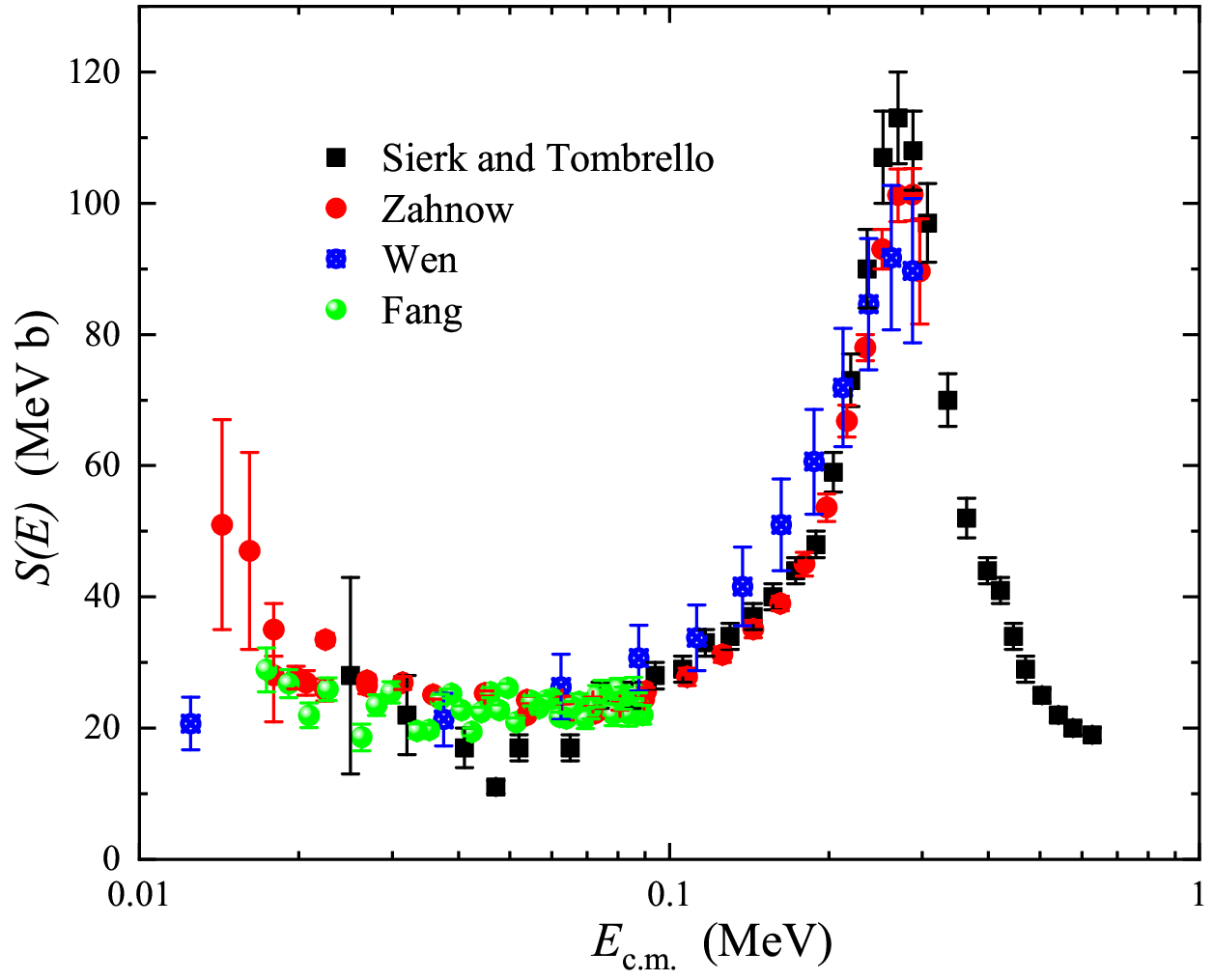}
\caption{(Color online) Astrophysical $S(E)$ factor for the ${}^{9}\text{Be}(p, \alpha){}^{6}\text{Li}$ reaction as a function of the center-of-mass energy ($E_{\text{c.m.}}$). The black squares represent the experimental data from Sierk and Tombrello \cite{Sierk1973}, while the red solid circles denote the results reported by Zahnow et al. \cite{Zahnow1997}. The blue solid squares correspond to the data extracted via the THM by Wen et al. \cite{Wen2008}, and the green solid circles highlight the measurements from Fang et al. \cite{Fang2018}.}
\label{wiii}
\end{figure}

The ${}^{9}\text{Be}(p, \alpha){}^{6}\text{Li}$ reaction is a means for beryllium destruction in astrophysical environments. Figures~\ref{wii} and \ref{wiii} present the measured reaction cross-sections and the corresponding astrophysical $S(E)$ factors as functions of the center-of-mass energy ($E_{\text{c.m.}}$), respectively.

In these figures, the experimental data are compiled from multiple high-precision measurements: the black squares represent the data reported by Sierk and Tombrello \cite{Sierk1973}, while the red solid circles denote the results from Zahnow et al. \cite{Zahnow1997}. Furthermore, the blue solid squares illustrate the bare nuclear data obtained via the Trojan Horse Method by Wen et al. \cite{Wen2008}, and the green solid circles highlight the most recent measurements from Fang et al. \cite{Fang2018}.

A critical comparison of these datasets reveals notable variations in both the magnitude and energy dependence of the reaction rate. These discrepancies likely stem from differences in experimental setups, background subtraction techniques, and data processing methodologies adopted by each research group. In the following sections, we will analyze these discrepancies in detail, focusing on how different analytical approaches—particularly the THM-based extraction and the interference-corrected Breit-Wigner fitting—influence the determination of the $S(E)$ factor at sub-barrier energies. This analysis is essential for mitigating the uncertainties associated with direct measurements and establishing a more robust benchmark for stellar evolution models \cite{Tumino2025, Lamia2015, Xu2013, Adelberger2011}.

\subsection{Analysis of direct measurement data: R-matrix versus interference-corrected Breit-Wigner approaches}

The experimental study by Sierk and Tombrello \cite{Sierk1973} provides a foundational dataset for the ${}^{9}\text{Be}(p, \alpha){}^{6}\text{Li}$ and ${}^{9}\text{Be}(p, d){}^{8}\text{Be}$ reactions. In their pioneering work, cross-sections were measured for proton laboratory energies ranging from 30 to 700 keV, with angular distributions obtained between 100 and 600 keV. A prominent resonance was observed at $E_{\text{lab}} = 330$ keV, identified as a $J^{\pi} = 1^{-}$ state in the ${}^{10}\text{B}$ compound nucleus. The peak cross-sections at this resonance were reported as $360 \pm 20$ mb for the $(p, \alpha)$ channel and $470 \pm 30$ mb for the $(p, d)$ channel, leading to an estimated zero-energy astrophysical $S(E)$ factor, $S_{\text{c.m.}}(0)$, of $35_{-15}^{+45}$ MeV$\cdot$b for the combined channels. Theoretical interpretation was subsequently performed using an R-matrix model, which incorporated three resonance levels in ${}^{10}\text{B}$: a $J^{\pi} = 2^{+}$ level at $-20$ keV, a $1^{-}$ level at $310$ keV, and a $1^{+}$ level at $410$ keV (all energies in c.m. frame). Sierk and Tombrello concluded that the reaction cross-sections in this region were dominated by compound nucleus formation, with negligible contributions from direct reaction mechanisms.

It is important to note that the significant role of the electron screening effect in low-energy nuclear reactions was not yet recognized during that era. Consequently, this effect was not incorporated into their analysis. Nevertheless, their work serves as a foundational reference for understanding the low-energy behavior of ${}^{9}\text{Be}+p$ reactions and remains essential for benchmarking modern theoretical models and astrophysical nucleosynthesis calculations \cite{Sierk1973}.

As experimental techniques evolved, the necessity to resolve discrepancies in the ultra-low energy region (below 100 keV) and to investigate novel physical effects became apparent. Building upon the framework established by Sierk and Tombrello, Zahnow et al. \cite{Zahnow1997} conducted refined measurements spanning the energy range of 16 to 390 keV. While their data exhibited general agreement with earlier results at overlapping energies, Zahnow et al. achieved higher signal-to-noise ratios, enabling the first observation of the electron screening effect in this reaction \cite{Greife1995, Aliotta2001, Engstler1992}. The significant discrepancy between the deduced screening potential ($U_{e} \approx 900$ eV) and the theoretical adiabatic limit ($\approx 240$ eV) remains a critical point of contention in current low-energy nuclear reaction studies.

\begin{table}[htbp]
    \centering
    \caption{Parameters of the ${}^{9}\text{Be}(p, \alpha){}^{6}\text{Li}$ reaction and screening potential. Data from Ref.~\cite{Fang2018,Zahnow1997,Sierk1973}.}
    \setlength{\tabcolsep}{1.8mm} \renewcommand\arraystretch{1.5}
    \label{tab:parameters}
    \begin{tabular}{lccc}
        \hline
        \hline
        Parameter & Fang & Zahnow & Sierk and Tombrello \\ 
        \midrule
        $E_{\text{R}}$ (keV) & $338 \pm 6$ & $336 \pm 3$ & $319 \pm 6$ \\
        $\Gamma_{\text{R}}$ & $194 \pm 8$ & $205 \pm 6$ & $133 \pm 6$ \\
        $\Gamma_{\alpha\text{R}}$ & $37 \pm 6$ & $68 \pm 2$ & $\approx 50$ \\
        $\Gamma_{\text{dR}}$ & $78 \pm 9$ & $90 \pm 4$ & $\approx 43$ \\
        $S_{\text{D}}$ (MeV) & $13.7 \pm 1.8$ & $14.2 \pm 0.5$ & --- \\
        $X_{\text{D}}$ (\%) & $6.8 \pm 0.1$ & $2.7 \pm 0.5$ & --- \\
        $U_{\text{s}}$ (eV) & $545 \pm 98$ & $900 \pm 50$ & --- \\
        $\chi^2/\text{ndf}$  & $2.9$ (76) & $1.5$ (33) & --- \\
        
        \hline
    \end{tabular}
    \label{I}
\end{table}

To address discrepancies in the sub-Coulomb barrier region, Fang et al. \cite{Fang2018} extended the experimental reach by measuring thick-target $\alpha$-yields in the ultra-low energy range of $E_{p} = 18\text{--}100$ keV. This dataset, combined with higher-energy results from Sierk and Tombrello \cite{Sierk1973} and Zahnow et al. \cite{Zahnow1997}, provides a robust basis for deducing the astrophysical $S(E)$ factor and the screening potential $U_{s}$.

Fang et al. employed both an interference-corrected Breit-Wigner method and a comprehensive multi-level R-matrix analysis using the AZURE2 code \cite{Lane1958, Barker2001}. The Breit-Wigner analysis yielded a screening potential of $U_{s} = 545 \pm 98$ eV, which remains significantly higher than the adiabatic limit of $\sim 264$ eV. To further refine the interpretation, a sophisticated level scheme of ${}^{10}\text{B}$ ($1^{-}, 2^{+}, 2^{-}, 1^{+}, 3^{+}$ states) was incorporated into the R-matrix framework.

The resulting R-matrix parameters, derived from the extracted level scheme, yield a zero-energy bare astrophysical $S(E)$ factor of $S_{\text{bare}}(0) = 17.4$ MeV$\cdot$b and a screening potential of $U_{s} = 538$ eV. The mutual agreement between the Breit--Wigner parametrization and the $R$-matrix analysis suggests that the $1^{-}$ resonant contribution is consistently described within both frameworks, supporting the physical reliability of the resonance parameters deduced from the direct measurement data.

However, the interpretation of direct measurement data is often constrained by the presence of the electron screening effect, which may mask the subtle signatures of sub-threshold resonances. Furthermore, the anomalously high screening potential obtained in these analyses indicates that current atomic physics models may be insufficient to fully describe the target environment. Consequently, Fang et al. \cite{Fang2018} concluded that future studies employing inverse kinematics are essential to definitively characterize the screening phenomenon and eliminate target-related uncertainties, thereby providing a more robust determination of the bare nuclear cross-section.

\subsection{Polynomial fitting, Breit-Wigner parametrization, and R-matrix analysis of THM data}

The extraction of astrophysical information from THM data necessitates a multi-faceted analytical approach. In contrast to direct measurements, which are often susceptible to electron screening effects at ultra-low energies, THM provides a unique probe for determining the bare nuclear cross-section.

By comparing the results derived from these complementary methodologies, we can effectively disentangle the complex interplay between compound nucleus formation and non-resonant direct reaction contributions. Furthermore, the integration of THM-based data—which inherently bypasses the Coulomb barrier—with R-matrix parameters refined from traditional direct measurements allows for a significant reduction in the uncertainties associated with the $S(0)$ factor extrapolation. This multi-methodological consistency check is essential for establishing a robust benchmark for nucleosynthesis models.

In Figure~\ref{wiii}, the THM data clearly reproduce the resonance peak observed in direct measurements near $E_{\text{c.m.}} = 270$ keV, which corresponds to the $J^{\pi}=1^{-}$ level of ${}^{10}\text{B}$ at an excitation energy of 6.87 MeV. While the THM data show good agreement with direct measurements above 100 keV, a significant divergence appears below this energy: the direct experimental data show a pronounced enhancement, which is widely attributed to the electron screening effect\cite{Barker2001,Zahnow1997,Fang2018}. Crucially, the THM approach is inherently free from such screening effects, as the reaction occurs at quasi-free conditions, providing a direct probe of the bare nuclear $S(E)$ factor.

\begin{figure}[tbp]
\centering
\includegraphics[width=10cm]{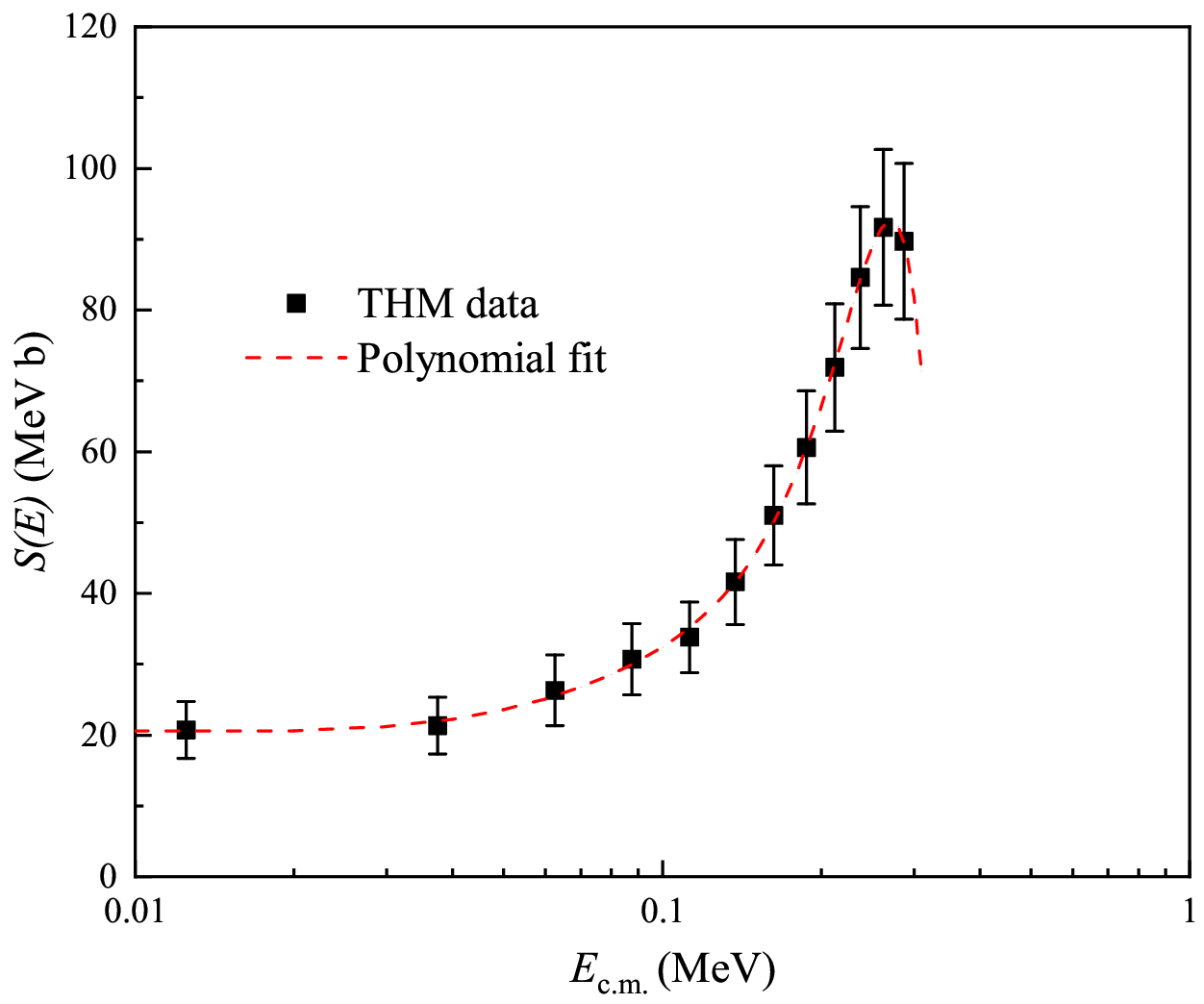}
\caption{(Color online) Astrophysical $S(E)$ factor for the ${}^{9}\text{Be}(p, \alpha){}^{6}\text{Li}$ reaction derived from THM data. The solid squares represent the experimental THM data, and the red dashed curve denotes the results of the fifth-order polynomial fit, with coefficients detailed in Table~\ref{II}.}
\label{wv}
\end{figure}

To derive the astrophysical $S(0)$ factor, Wen et al. \cite{Wen2008} performed a fifth-order polynomial fit to the THM data, given by the expression:

\begin{equation}
\begin{split}
    S(E) = & S(0) + S_{1}E + S_{2}E^{2} + S_{3}E^{3} \\
           & + S_{4}E^{4} + S_{5}E^{5}
\end{split}
\label{eq:poly_fit}
\end{equation}

The fitting coefficients are detailed in Table~\ref{II}, and the resulting fit is illustrated by the solid curve in Figure~\ref{wv}. Based on this analysis, the astrophysical $S(E)$ factor at zero energy was determined to be $S(0) = 21.0 \pm 0.8$ MeV$\cdot$b. This value provides a critical benchmark for evaluating the bare reaction rate, free from the complexities of the environmental electron screening encountered in conventional target-based experiments.

\begin{table}[htbp]
    \centering
    \caption{Coefficients of the fifth-order polynomial fit for the $S(E)$ factor.}
    \setlength{\tabcolsep}{4.2mm}\renewcommand\arraystretch{1.5}
    \label{tab:poly_fit}
    \begin{tabular}{lcc}
    \hline
    \hline        
        Coefficients & Value & Error \\ 
        \midrule
        $S(0)$ (MeV b) & $21.0$ & $\pm 0.8$ \\
        $S_1$ (b) & $-92.4$ & $\pm 13.4$ \\
        $S_2$ (MeV$^{-1}$ b) & $4669$ & $\pm 78.5$ \\
        $S_3$ (MeV$^{-2}$ b) & $-4.413 \times 10^4$ & $\pm 331$ \\
        $S_4$ (MeV$^{-3}$ b) & $2.193 \times 10^5$ & $\pm 1251$ \\
        $S_5$ (MeV$^{-4}$ b) & $-3.768 \times 10^5$ & $\pm 3483$ \\
        $\chi^2/\text{ndf}$ & $5.3/6$ & --- \\ 
        \hline
    \end{tabular}
    \label{II}
\end{table}

\begin{figure}[tbp]
\centering
\includegraphics[width=10cm]{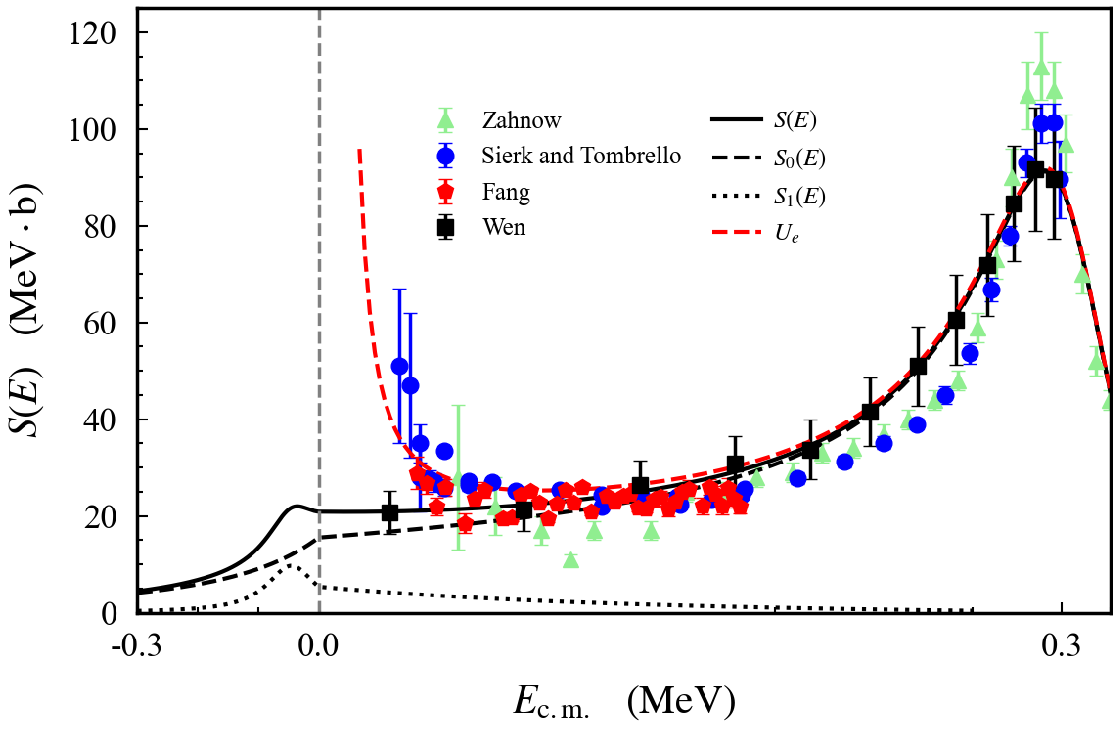}
\caption{(Color online) Astrophysical $S(E)$ factor results obtained from Breit-Wigner function fitted THM data compared with direct measurement fitted data.}
\label{wiv}
\end{figure}

To account for both the sub-threshold and resonant structures, the THM data are parameterized using a double Breit-Wigner function. We employ a dual-resonance model composed of $S_{0}(E)$ and $S_{1}(E)$, where $S_{0}(E)$ describes the dominant resonance peak above the threshold, while $S_{1}(E)$ accounts for the sub-threshold resonance located at $E_{1} = -23$ keV.

As shown in Figure~\ref{wiv}, the THM-derived $S(0)$ value is notably higher than the values extrapolated from traditional direct measurements. In this figure, the red long-dashed curve represents the fit to direct measurement data accounting for the electron screening effect, while the black solid curve displays the THM data fitted using the double Breit-Wigner function, which accounts for both the above-threshold and sub-threshold resonance peaks. Furthermore, the black long-dashed curve and the black dotted curve illustrate the above-threshold resonance term $S_{0}(E)$ and the sub-threshold resonance term $S_{1}(E)$, respectively.

This discrepancy is largely attributed to the contribution of the sub-threshold resonance structure. In our analysis, we distinguish between two extrapolations: $S_{0}(0)$, calculated based solely on the primary resonance, and $S(0)$, the total astrophysical $S(E)$ factor derived by incorporating the sub-threshold resonance contribution within the dual-resonance model. The refined fitting parameters are summarized in Table~\ref{III} ($\chi^2/\text{ndf} \approx 0.0437$).

The BW analysis demonstrates that the inclusion of the sub-threshold resonance improves the description of the experimental data trend in the low-energy region. This suggests that the contribution of the sub-threshold state is essential for reconciling the THM results with direct measurements, thereby providing a more robust extrapolation toward the stellar energy regime.

\begin{table}[htbp]
\caption{Fitting parameters for the THM data derived using the double Breit-Wigner function.}
\label{tab:bw_parameters}
\[
\setlength{\tabcolsep}{2.35mm}\renewcommand\arraystretch{2}
\begin{tabular}{ccccccccc}
\hline\hline
A$_{0}$ & E$_{0}$/keV & $\Gamma_{0}$/keV & A$_{1}$ & E$_{1}$/keV & $\Gamma_{1}$/keV & S$_{0}$(0)/MeV$\cdot$b & S(0)/MeV$\cdot$b \\ \hline
92$\pm$8 & 279$\pm$33 & 260$\pm$42 & 10$\pm$4 & $-23$ & 48$\pm$73 & 16.3 & 21.5 \\ 
\hline
\end{tabular}
\]
\label{III}
\end{table}

\begin{table}[htbp]
\caption{Extracted $R$-matrix parameters for the $J^{\pi}=1^{-}$ and $2^{+}$ states. Results above the double horizontal line correspond to the fit primarily associated with the $1^{-}$ resonance, while those below the line correspond to the fit including the dominant contributions from both the $1^{-}$ resonance and the sub-threshold $2^{+}$ state.}
\label{tab:r_matrix_combined}
\small
\[
\setlength{\tabcolsep}{4.3pt}
\renewcommand\arraystretch{1.5}
\begin{tabular}{ccccccccccc}
\hline
\hline 
 & & & \multicolumn{2}{c}{$\gamma_{\lambda ps} (\text{MeV}^{1/2})$} & \multicolumn{3}{c}{$\gamma_{\lambda \alpha l'} (\text{MeV}^{1/2})$} & \multicolumn{3}{c}{$\gamma_{\lambda dl'} (\text{MeV}^{1/2})$} \\
\cmidrule(lr){4-5} \cmidrule(lr){6-8} \cmidrule(lr){9-11}
$J^{\pi}$ & $\lambda$ & $E_{\lambda}$ (MeV) & $s=1$ & $s=2$ & $l'=0$ & $l'=1$ & $l'=2$ & $l'=0$ & $l'=1$ & $l'=2$ \\ 
\midrule
$1^-$ & 1 & 6.884 & $0.640$ & --- & --- & $0.357$ & --- & --- & 0.00271 & --- \\\hline\hline
$2^+$ & 1 & 6.560 & $-0.361$ & $0.476$ & $0.244$ & $2.233$ & --- & --- & --- & --- \\
$1^-$ & 1 & 6.888 & $0.654$ & --- & --- & $0.363$ & --- & --- & 0.000155 & --- \\ 
\hline
\end{tabular}
\]
\label{IV}
\end{table}

\begin{figure}[tbp]
\centering
\includegraphics[width=10cm]{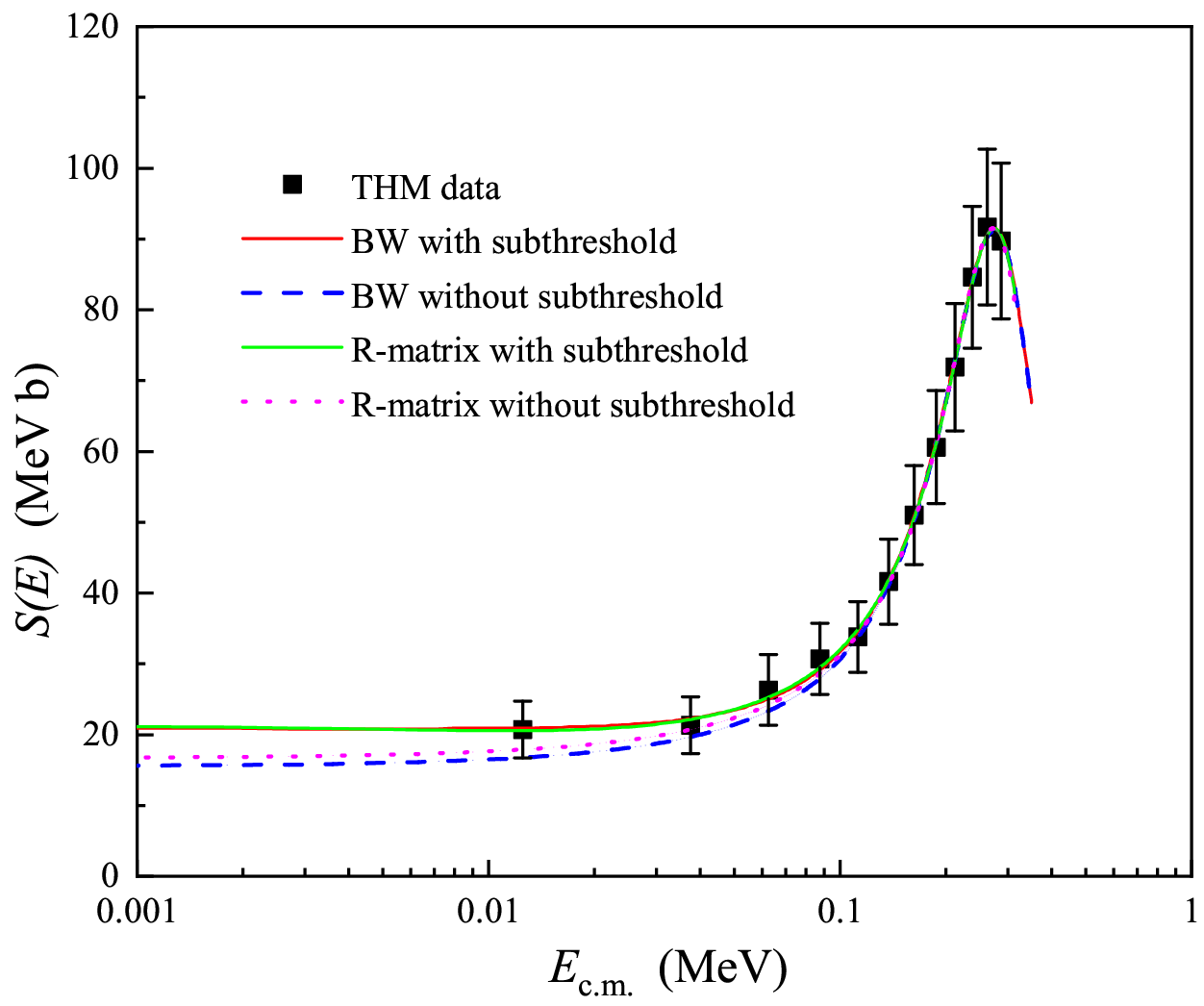}
\caption{(Color online) Astrophysical $S(E)$ factor for the ${}^{9}\text{Be}(p, \alpha){}^{6}\text{Li}$ reaction. The solid squares represent the experimental data extracted via the THM. The curves correspond to different fitting approaches: the solid red and dashed blue lines denote the Breit-Wigner (BW) fits with and without the sub-threshold resonance, respectively, while the green solid and magenta dotted lines represent the R-matrix fits with and without the sub-threshold resonance, respectively.}
\label{wvi}
\end{figure}

To further investigate the low-energy characteristics of the ${}^{9}\text{Be}(p, \alpha){}^{6}\text{Li}$ reaction, we performed a systematic analysis of the THM data within the R-matrix theoretical framework. In all fits, we adopted the conventional values for the channel radii: $a_c = 1.45(A_1^{1/3} + A_2^{1/3})$ fm, yielding $a_p = 4.466$ fm, $a_{\alpha} = 4.937$ fm, and $a_d = 4.727$ fm. The boundary conditions ($B_c$ values) were chosen to be equal to the shift factor at a representative excitation energy for each $J^{\pi}$ value \cite{Lane1958, Sierk1973,Zahnow1997,Barker2001,Fang2018}.

The $R$-matrix parameters obtained from the multi-channel fit are summarized in Table~\ref{IV}. The parameters listed above the double horizontal line correspond to the fit dominated by the $1^{-}$ resonance, whereas those listed below the line correspond to the fit in which both the $1^{-}$ and $2^{+}$ resonances provide the dominant contributions.

As shown in Figure~\ref{wvi}, the model is in good agreement with the THM experimental data, particularly in the region of the $1^{-}$ resonance. A primary objective of this analysis is to quantify the impact of the sub-threshold resonance on the astrophysical $S(0)$ factor. The results summarized in Table~\ref{IV} demonstrate that the inclusion of the sub-threshold resonance significantly alters the extrapolated $S(0)$ factor. Specifically, without accounting for the sub-threshold resonance, we obtain $S(0) = 16.7$ MeV$\cdot$b ($\chi^2/\text{ndf} \approx 0.006$). Conversely, upon incorporating the sub-threshold resonance effect, the R-matrix model yields $S(0) \approx 21.1$ MeV$\cdot$b ($\chi^2/\text{ndf} \approx 0.022$).

This increase is consistent with the energy dependence observed in the THM data at ultra-low energies. The agreement between the double Breit--Wigner and $R$-matrix descriptions provides strong support for the interpretation that a sub-threshold state at approximately $-23$ keV is likely responsible for a substantial part of the observed enhancement of the astrophysical $S(E)$ factor in the low-energy region. Furthermore, compared with previous extrapolations based exclusively on direct measurements \cite{Sierk1973,Fang2018}, the inclusion of the sub-threshold contribution yields a more accurate description of the bare-nucleus $S(E)$ factor and provides a consistent theoretical interpretation over the entire energy range. These findings highlight the importance of sub-threshold effects in determining the low-energy behavior of the $^{9}\mathrm{Be}(p,\alpha)^{6}\mathrm{Li}$ reaction.

\section{Reaction Rate and Astrophysical Implications}

\subsection{Reaction Rate}

The surface abundance of ${}^{9}\text{Be}$ in stars is significantly influenced by key physics inputs, such as the astrophysical reaction rate, the equation of state, the opacity of the stellar matter, and external convection efficiencies. The reaction rate at astrophysical energies has been deduced using the standard formula provided by Rolfs \& Rodney (1988) \cite{Rolfs1988}:

\begin{equation}
\begin{split}
N_A \langle \sigma v \rangle = & \left( \frac{8}{\pi \mu} \right)^{1/2} \frac{N_A}{(k_B T_9)^{3/2}} \\
& \times \int_0^\infty S_b(E) \exp\left( -2\pi\eta - \frac{E}{k_B T_9} \right) dE \quad
\end{split}
\end{equation}

where the temperature $T_9$ is expressed in units of $10^9$ K, and the center-of-mass energy $E$ is in units of MeV. In Equation (11), $S_b(E)$ represents the bare-nucleus astrophysical $S(E)$ factor. The integration is performed across the energy intervals covered by the experimental data, spanning from approximately 200 keV down to 10 keV, depending on the specific reaction channel.

\begin{figure}[tbp]
\centering
\includegraphics[width=10cm]{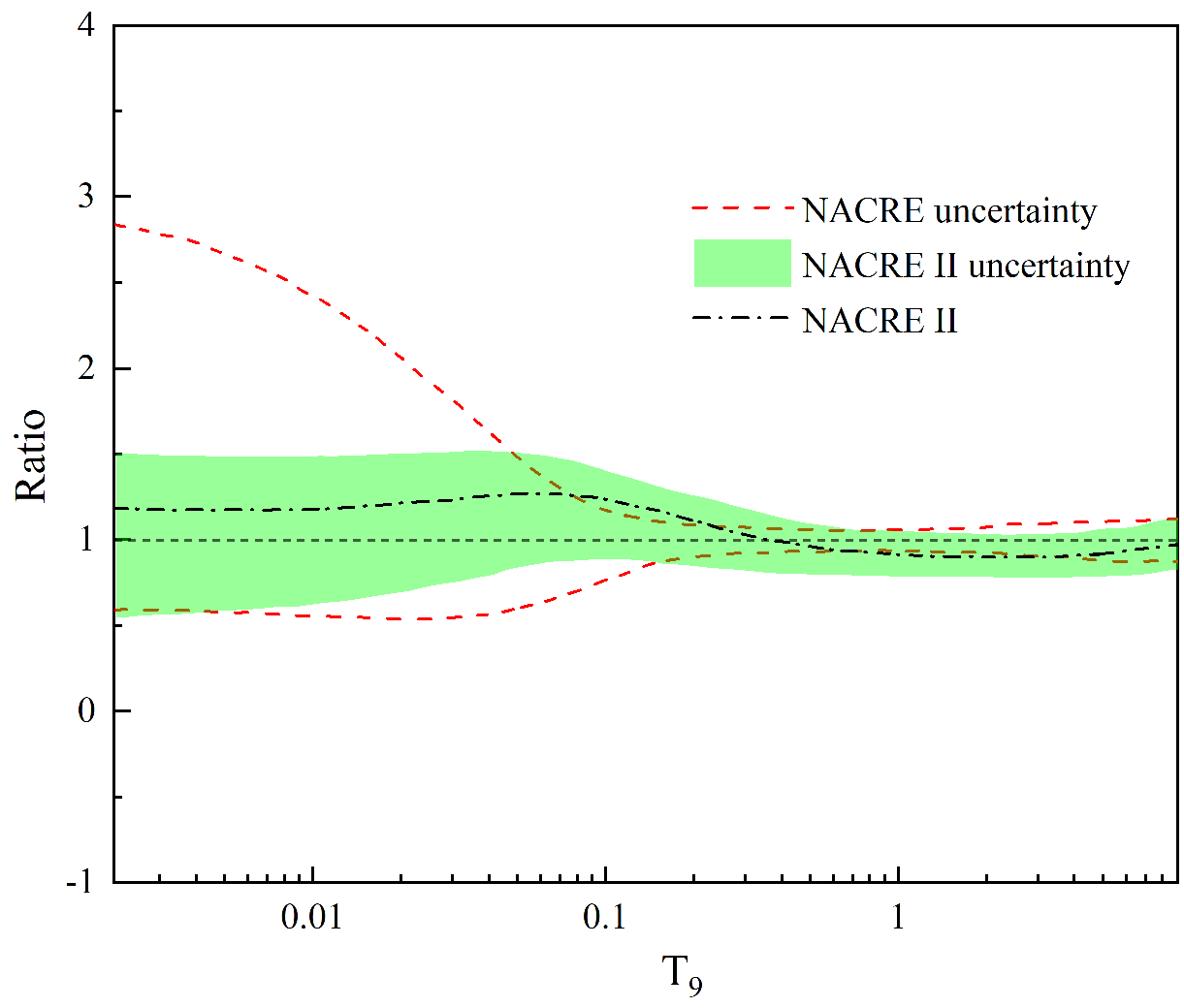}
\caption{(Color online) $^9\mathrm{Be}(\mathrm{p},\alpha)^6\mathrm{Li}$ rates in units of the NACRE (adopt) values. Data from Ref.~\cite{Xu2013}.}
\label{wvii}
\end{figure}

In the NACRE II compilation \cite{Xu2013}, the astrophysical $S(E)$ factor for the ${}^{9}\text{Be}(p, \alpha){}^{6}\text{Li}$ reaction was systematically re-evaluated by integrating historical direct measurements with post-NACRE experimental results, as shown in Figure~\ref{wvii}. A pivotal advancement in NACRE II was the incorporation of THM data \cite{Wen2008}, which extended the experimental energy range down to $E_{\text{c.m.}} \simeq 12$ keV. By comparing these THM results with direct measurements, Xu et al. \cite{Xu2013} demonstrated that the cross-section enhancements observed in traditional datasets below $E_{\text{c.m.}} \simeq 30\text{--}40$ keV are predominantly attributable to electron screening rather than intrinsic nuclear properties. Consequently, these low-energy direct data were excluded from their DWBA-based potential model fit to isolate the bare nuclear cross-section, yielding a zero-energy $S$-factor of $S(0) = 21^{+5}_{-13}$ MeV$\cdot$b, which significantly updated the previous NACRE value of $17^{+25}_{-7}$ MeV$\cdot$b.

\begin{figure}[tbp]
\centering
\includegraphics[width=10cm]{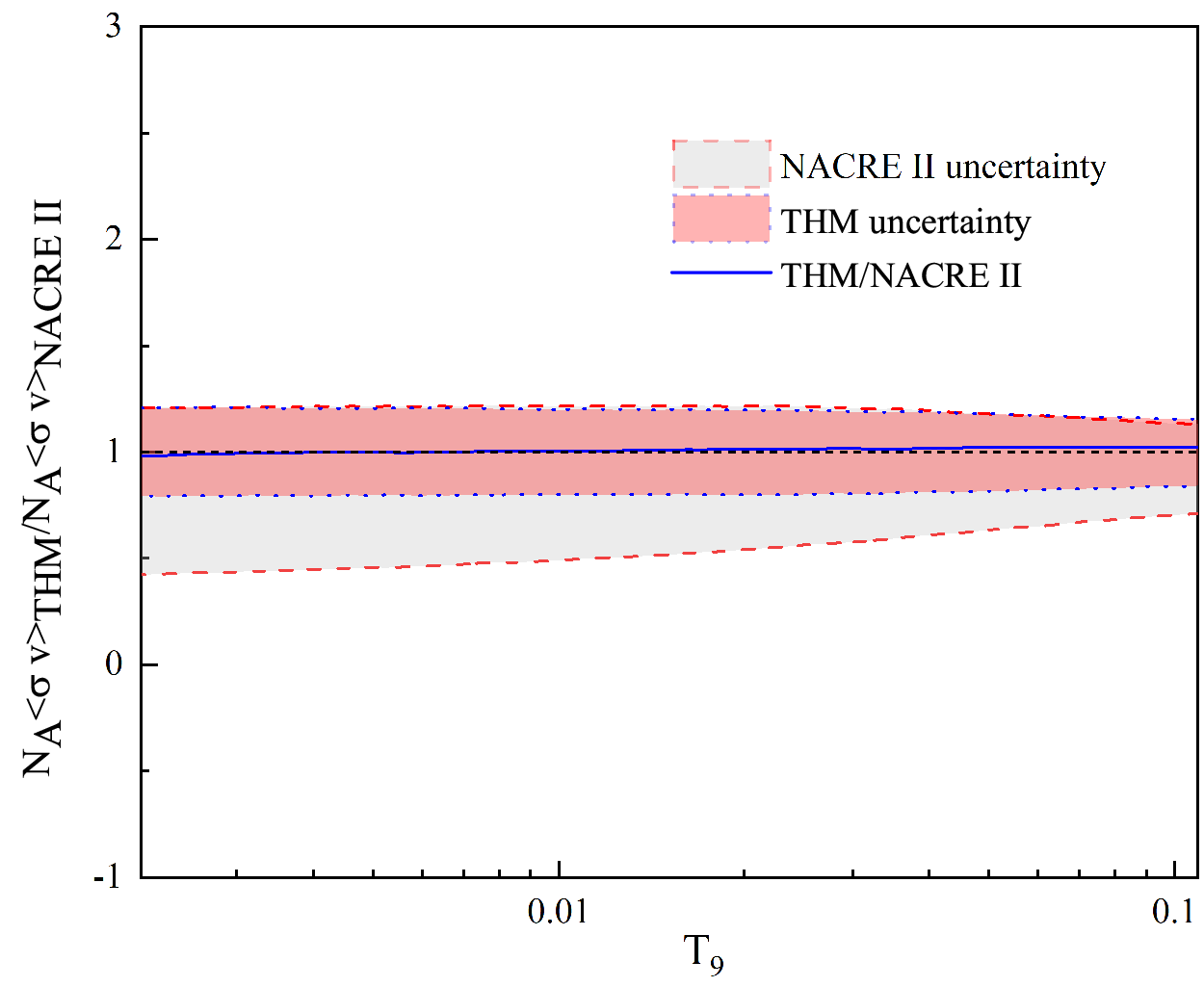}
\caption{(Color online) Ratio of the ${}^{9}\text{Be}(p,\alpha){}^{6}\text{Li}$ reaction rate to the NACRE II compilation. The blue curve represents the ratio of the THM-derived reaction rate to the NACRE II evaluation. Data from Ref.~\cite{Lamia2015}.}
\label{wviii}
\end{figure}

Lamia et al. \cite{Lamia2015} carried out a systematic investigation of astrophysical reaction rates extracted from the THM and compared them with the recommended values from the second-generation Nuclear Astrophysics Compilation of Reaction Rates (NACRE II), as shown in Figure~\ref{wviii}.

The ratio of the THM-derived $^{9}\mathrm{Be}(p,\alpha)^{6}\mathrm{Li}$ reaction rate to the NACRE II evaluation remains nearly constant throughout the investigated temperature range, demonstrating the overall consistency between the two evaluations. In the astrophysically relevant low-temperature region ($T<10^{8}$ K), the uncertainty of the THM-based reaction rate is reduced to about $20\%$, compared with $70\%$--$90\%$ for the NACRE II evaluation.

Furthermore, the uncertainty band associated with the THM-based reaction rate (red shaded region) is significantly narrower than that of NACRE II (gray shaded region), highlighting the important role of THM measurements in constraining the low-energy astrophysical $S(E)$ factor and improving the reliability of stellar reaction-rate calculations.

Such agreement is not surprising, since the THM measurements were already incorporated by Xu et al. \cite{Xu2013} as one of the main experimental inputs for determining the low-energy astrophysical $S(E)$ factor.

Building upon the available THM measurements, the present work employs both the $R$-matrix approach and a double Breit--Wigner formalism to analyze the low-energy astrophysical $S(E)$ factor. The comparison between the two descriptions provides independent evidence for the significant contribution of the sub-threshold resonance to the low-energy enhancement of the astrophysical $S(E)$ factor. The consistency between the two approaches strengthens the physical interpretation of the THM data and provides a more reliable determination of the low-energy extrapolation of the astrophysical $S(E)$ factor.

These refined theoretical analyses enhance our understanding of the underlying reaction dynamics and provide a more robust basis for calculating stellar reaction rates relevant to nuclear astrophysics.

\subsection{Astrophysical Implications}

Stellar evolution studies by Lamia et al. have shown that, although the burning of ${}^{9}\mathrm{Be}$ contributes negligibly to the overall stellar energy budget and therefore does not affect the global stellar structure or evolutionary track, it can significantly influence the surface abundance of ${}^{9}\mathrm{Be}$.

During stellar evolution, the surface abundance of ${}^{9}\mathrm{Be}$ is highly sensitive to the burning efficiency at the base of the convective envelope, which in turn depends directly on the accuracy of the low-energy reaction cross section and reaction rate. Consequently, reducing the uncertainties associated with the relevant nuclear-physics inputs is of considerable importance for improving the reliability of light-element abundance calculations.

The bare-nucleus astrophysical factor obtained from THM measurements effectively removes the influence of electron screening and substantially reduces the uncertainty of the reaction rate in the low-temperature region. As a result, it provides a more reliable nuclear-physics input for stellar evolution models. Previous studies have shown that the THM-based ${}^{9}\mathrm{Be}(p,\alpha){}^{6}\mathrm{Li}$ reaction rate is approximately $25\%$ higher than the NACRE recommended value, leading to a faster depletion of ${}^{9}\mathrm{Be}$ and consequently lower surface Be abundances at a given stellar age. For higher-mass stars ($M>0.5,M_{\odot}$), the abundance variation induced by the updated reaction rate is relatively small. In contrast, for low-mass pre-main-sequence stars ($M\le0.5,M_{\odot}$), where Be burning occurs at temperatures of several million Kelvin, the effect becomes considerably more pronounced. In some low-effective-temperature regions ($T_{\mathrm{eff}}\lesssim3600,\mathrm{K}$), the difference in the logarithmic ${}^{9}\mathrm{Be}$ abundance predicted using the THM and NACRE reaction rates can reach approximately 1 dex.

Since Be abundances have long been regarded as sensitive probes of internal mixing, convection, and transport processes in stellar interiors, a more accurate determination of the ${}^{9}\mathrm{Be}(p,\alpha){}^{6}\mathrm{Li}$ reaction rate is expected to reduce the nuclear-physics uncertainties in light-element abundance predictions and provide a more reliable basis for constraining stellar structure and evolutionary processes through future Be-abundance observations.

\section{Summary and Outlook}

In this work, we have systematically investigated the low-energy behavior of the ${}^{9}\mathrm{Be}(p,\alpha){}^{6}\mathrm{Li}$ reaction and its astrophysical $S(E)$ factor. By combining high-precision direct measurement data \cite{Sierk1973,Zahnow1997,Fang2018} with bare-nucleus benchmark data obtained using the Trojan Horse Method (THM) \cite{Wen2008}, we performed a comprehensive study of the underlying reaction mechanisms within the stellar energy region.

The comparative analysis indicates that the cross-section enhancement observed in direct measurements below 100 keV is predominantly caused by electron-screening effects. In contrast, the THM data, being free from such environmental influences, provide an essential constraint on the bare-nucleus astrophysical $S(E)$ factor. To investigate the low-energy behavior of the reaction, a multi-model framework based on polynomial fitting, a double Breit--Wigner (BW) formalism, and a multi-level $R$-matrix analysis implemented with the \textsc{AZURE2} code was employed. 

The results show that the inclusion of a sub-threshold resonance substantially improves the description of the experimental trend observed in the THM data. In particular, the introduction of a sub-threshold state at $E_{1}=-23$ keV leads to a zero-energy bare-nucleus astrophysical factor of $S_{\mathrm{bare}}(0)\approx 21.0$--$21.5$ MeV$\cdot$b, significantly larger than the values obtained from single-level approximations. The agreement between the double Breit--Wigner and $R$-matrix descriptions provides strong support for the interpretation that a sub-threshold state at approximately $-23$ keV is likely responsible for a substantial part of the observed low-energy enhancement of the astrophysical $S(E)$ factor. Furthermore, the inclusion of the sub-threshold contribution enables a more accurate and self-consistent description of the bare-nucleus $S(E)$ factor over the entire energy range covered by the available experimental data.

The THM data have already been incorporated into the NACRE II reaction-rate compilation, leading to improved accuracy in astrophysical reaction-rate evaluations. Building upon these experimental constraints, the refined theoretical analyses presented in this work provide further insight into the underlying reaction dynamics and establish a more reliable foundation for stellar reaction-rate calculations. The results obtained here offer valuable guidance for future high-precision experimental investigations and for the development of theoretical models in nuclear astrophysics.

Future high-precision measurements in the ultra-low-energy region are required to clarify the origin of the so-called anomalous screening effect and to further constrain the properties of the proposed sub-threshold state. On the theoretical side, more sophisticated $R$-matrix analyses incorporating improved spectroscopic information on the ${}^{10}\mathrm{B}$ compound nucleus will be essential for reducing uncertainties in stellar nucleosynthesis calculations and for providing a more robust basis for light-element abundance predictions in the early universe.

%% Please use the acknowledgment and contribution environments. This will
%% be anonymized when the "anonymous" style option is used.
\begin{acknowledgments}
The authors gratefully acknowledge Prof. Shu-Hua Zhou from the China Institute of Atomic Energy for his invaluable guidance and insightful advice throughout this work. We also express our sincere gratitude to the Nuclear Reaction Group at the China Institute of Atomic Energy for their kind assistance. Additionally, the authors thank Dr. Yu Li from Hefei University of Technology for her constructive and helpful discussions.

This work was supported by the National Natural Science Foundation of China (Grant Nos. 12075031, 12275360, 12475115, and 11935001), the Natural Science Foundation of Beijing Municipality (Grant No. 1222022), and the Anhui Project (Grant No. Z010118169).

\end{acknowledgments}

\end{document}